\documentclass[conference]{IEEEtran}
\IEEEoverridecommandlockouts
\usepackage{cite}
\usepackage{amsmath,amssymb,amsfonts}
\usepackage{algorithmic}
\usepackage{graphicx}
\usepackage{textcomp}
\usepackage{xcolor}
\usepackage{booktabs}
\usepackage{graphbox}
\usepackage{color}
\usepackage{url}
\usepackage{multirow}
\usepackage{subcaption}
\usepackage{longtable}
\usepackage{colortbl}
\definecolor{mygray}{gray}{.9}
\definecolor{LightCyan}{rgb}{0.88,1,1}

\newcommand{\nop}[1]{}

\def \eg {\emph{e.g.}, }

\def\BibTeX{{\rm B\kern-.05em{\sc i\kern-.025em b}\kern-.08em
    T\kern-.1667em\lower.7ex\hbox{E}\kern-.125emX}}
\begin{document}

\title{Influence Pathway Discovery on Social Media
\thanks{This work was supported in part by DARPA award
HR001121C0165 and HR00112290105, and in part under
DoD Basic Research Office award HQ00342110002. It is also
supported in part by ACE, one of the seven centers in JUMP
2.0, a Semiconductor Research Corporation (SRC) program
sponsored by DARPA.}
}

\author{\IEEEauthorblockN{Xinyi Liu\textsuperscript{\textsection}, Ruijie Wang\textsuperscript{\textsection}, Dachun Sun\textsuperscript{\textsection}, Jinning Li, Christina Youn, You Lyu, Jianyuan Zhan, Dayou Wu, \\ 
Xinhe Xu, Mingjun Liu,  Xinshuo Lei, Zhihao Xu, Yutong Zhang, Zehao Li, Qikai Yang, Tarek Abdelzaher}
\IEEEauthorblockA{\textit{Department of Computer Science} \\
\textit{University of Illinois at Urbana-Champaign}\\
Urbana, IL 61801\\
\{liu323, ruijiew2, dsun18, jinning4, youn13, youlyu2, zhan32, dayouwu2, xinhexu2, \\ mingjun6, xinshuo3, zhihaox3, yutongz7, zehao3, qikaiy2, zaher\}@illinois.edu}
}

\maketitle

\maketitle
\begingroup\renewcommand\thefootnote{\textsection}
\footnotetext{Equal contributions.}
\endgroup

\begin{abstract}
This paper addresses {\em influence pathway discovery\/}, a key emerging problem in today's online media. We propose a discovery algorithm that leverages recently published work on {\em unsupervised interpretable ideological embedding\/}, a mapping of ideological beliefs (done in a self-supervised fashion) into {\em interpretable\/} low-dimensional spaces. Computing the ideological embedding at scale allows one to analyze correlations between the ideological positions of leaders, influencers, news portals, or population segments, deriving potential influence pathways. The work is motivated by the importance of
social media as the preeminent means for global interactions and collaborations on today's Internet, as well as their frequent (mis-)use to wield influence that targets social beliefs and attitudes of selected populations. Tools that enable the understanding and mapping of influence propagation through population segments on social media are therefore increasingly important. In this paper, influence is measured by the perceived ideological shift over time that is correlated with influencers' activity. Correlated shifts in ideological embeddings indicate changes, such as swings/switching (among competing ideologies), polarization (depletion of neutral ideological positions),   escalation/radicalization (shifts to more extreme versions of the ideology), or unification/cooldown (shifts towards more neutral stances). Case-studies are presented to explore selected influence pathways (i) in a recent French election, (ii) during political discussions in the Philippines, and (iii) for some Russian messaging during the Russia/Ukraine conflict.  
\end{abstract}


\section{Introduction}
The paper advances the science of {\em influence network tomography\/} - the empirical mapping of influence pathways on social media, derived from observable node behaviors. The work is motivated by the exploding popularity of 
social networks and their growing impact on population beliefs and attitudes. The paper posits that {\em interpretable unsupervised ideological embedding\/}~\cite{li2022unsupervised}, a recently proposed belief embedding approach, is a crucial enabler towards inferring potential influence among a diverse range of node types. Being unsupervised, it can be computed at scale for many parties with no need for human labeling. Being interpretable, it yields actionable insights. Since influence produces correlated changes in belief, computing the interpretable unsupervised ideological embedding at scale can serve as a new foundation for influence network tomography. As a proof of concept, an end-to-end self-supervised solution is developed that relies on uncovering correlated changes in nodes' ideological embeddings to discover potential influence pathways. Example outputs are presented based on multiple social media datasets that illustrate the versatility of the approach.

Modern solutions for influence estimation on social networks generally rely on a combination of (i) node embeddings (that distill node attributes correlated with predisposition to exert/follow influence), and (ii)  neural networks that learn to estimate influence among nodes given their computed embeddings. A pioneering paper in that space is DeepInf~\cite{qiu2018deepinf}, but many variations exist~\cite{leung2019personalized, liu2022modeling, cuzzocrea2020combined, xu2022mvinf, gao2022hetinf}. The ideological embedding we leverage~\cite{li2022unsupervised}, contrary to the aforementioned literature, is an {\em interpretable\/} node representation (computed using a self-supervised representation learning approach) that summarizes ideological beliefs. Changes in ideological embedding therefore directly denote interpretable phenomena such as polarization, radicalization, reconciliation, or desertion of ideological positions. Models that describe how changes in the embedding of one node affect another can therefore serve as interpretable indicators of ideological influence.


The rest of this paper is organized as follows. Section~\ref{sec:related} briefly covers key elements of related work. Section~\ref{sec:problem} presents the problem definition of {\em influence network tomography\/}. Section~\ref{sec:solution} describes the proposed approach for mapping influence pathways among diverse node types. Example case studies are covered in Section~\ref{sec:evaluation} based on several social media datasets. Section~\ref{sec:discussion}  describes insights gained from the evaluation and discusses opportunities for future extensions. The paper concludes with Section~\ref{sec:conclusion}. 

\section{Related Work}
\label{sec:related}
The work described in this paper falls in the broad category of influence estimation on social media. Early work on influence analysis in social networks generally computed influence either directly for individual network edges (resulting in such measures as tie strength~\cite{granovetter1973strength} and edge betweenness~\cite{freeman1977set}) or independently for individual nodes (resulting in such measures as node centrality~\cite{bonacich1972factoring} and closeness~\cite{freeman2002centrality}).
Early work also concerned itself with the analysis of epidemic diffusion cascades,  producing stylized models that predict cascade propagation, such as the linear threshold model~\cite{granovetter1978threshold} and the independent cascade model~\cite{goldenberg2001talk}. A large number of subsequent models were developed that differ in how diffusion probabilities were computed~\cite{guille2013information}. More recently, such models were adapted to utilize node embeddings and neural networks for diffusion prediction~\cite{bourigault2016representation,wang2021dydiff}. 

Influence estimation can be thought of as a generalization of diffusion prediction with the idea that information diffusion from one node to another is {\em one\/} manifestation of influence.
Modern influence estimation literature has focused on different flavors of node embeddings, where network nodes are mapped into latent spaces in a manner that allows predicting influence relations from such mappings. An example is Inf2vec~\cite{feng2018inf2vec}, which used relations among node embeddings to infer influence, and Multi-Influor~\cite{wang2020learning} that extended the approach to multiple influence factors from which influence could be predicted along network edges. Node embeddings are powerful because they can, in general, capture pairwise node interactions, local network structures, and global similarity across node categories. Furthermore, they allow the development of neural networks that take such embeddings as input and predict influence along edges as output. A well-cited early example of such an approach to influence estimation is DeepInf~\cite{qiu2018deepinf}.
It inspired many subsequent variants to use node embeddings and deep learning for influence prediction, differing in the details of the embedding used to capture node attributes and the subsequent neural network used to infer influence (e.g., see~\cite{leung2019personalized, liu2022modeling, cuzzocrea2020combined, xu2022mvinf, gao2022hetinf}).

In the fast-changing world of social media, it was also recognized that the underlying networks and attention patterns are not static. Influence relations are thus very time-dependent (in addition to being topic-dependent). Recent research therefore introduced time-sensitive
and topic-specific solutions for influence measurement~\cite{zheng2020measuring} and diffusion prediction~\cite{sankar2020inf}. Our approach is an instance of this latter category of techniques.

As alluded to in the introduction, our approach is novel in its reliance on a new type of node embedding that we call {\em interpretable unsupervised ideological embedding\/}~\cite{li2022unsupervised}. As its name suggests, the novelty of that embedding comes from being both interpretable and unsupervised -- a combination that had not been previously accomplished jointly in ideological mapping solutions. The embedding projects nodes into an ideological space, where different axes denote different ideologies (automatically disentangled in a self-supervised fashion from observing social media posts). On each axis, a positioning further up the axis denotes adherence to a more extreme version of the underlying ideology. A change in a node's embedding represents a change in the node's beliefs. The approach projects both {\em users\/} and {\em content\/} into the {\em same\/} latent space. Thus, the embedding (and subsequently influence propagation) can be computed for a more diverse set of node types, including individuals, communities, articles, and news portals. Furthermore, the projection into the ideological space is independent of the underlying social network platform; the embedding can be computed for users and posts on different social media platforms regardless of their format, such as text~\cite{polarization,CMH,li2022unsupervised} and images~\cite{liu2023unsupervised}. Below, we present the problem definition in more detail and then describe the solution architecture.

\section{Problem Definition}
\label{sec:problem}
Given posts on some set of social media, the purpose of this work is to uncover the latent {\em influence network\/}, comprised of {\it entities\/} and {\it directional edges}, that summarize how influence propagates on the media considered. In this network, the entities refer to objects in either the physical world or online that can influence one another. Each directional edge represents influence exerted by one entity on another. To set the scope, we consider four types of entities in analyzing influence pathways: (i) {\em physical entities\/}, such as events that transpire in the physical world, external to the social medium (such as protests, hospitalizations, donations, or deaths), (ii) {\em individual influencers\/}, whose positions allow them to impact larger population segments, (iii) {\em user community clusters\/} representing communities on social platforms defined by a clustering algorithm or according to the operator, and (iv) {\em information domains\/}, each associated with a specific agency portal (\eg YouTube, individual News media, etc) that publish information of a given type, such as {\tt cnn.com\/}, {\tt foxnews.com\/}, or {\tt infowars.com\/}. Of the above types of entities, the first type is represented by a directly measured time-series. For example, we can use the GDELT database\footnote{https://www.GDELTproject.org/} to count physical events of a particular type over time. 
Broadly speaking, physical entities can denote any real-valued time-series measurement to correlate other nodes' embeddings with. For example, economic indicators such as the price of crude oil or the consumer confidence index could constitute valid entities to consider in the influence network (e.g., to help understand the impact of social media activity on such indicators or vice versa).
The remaining three types of entities are represented by a time-series of ideological embedding versus time, computed as we describe later.
 
\begin{figure}[htp]
    \centering
    \includegraphics[width = 1.\columnwidth]{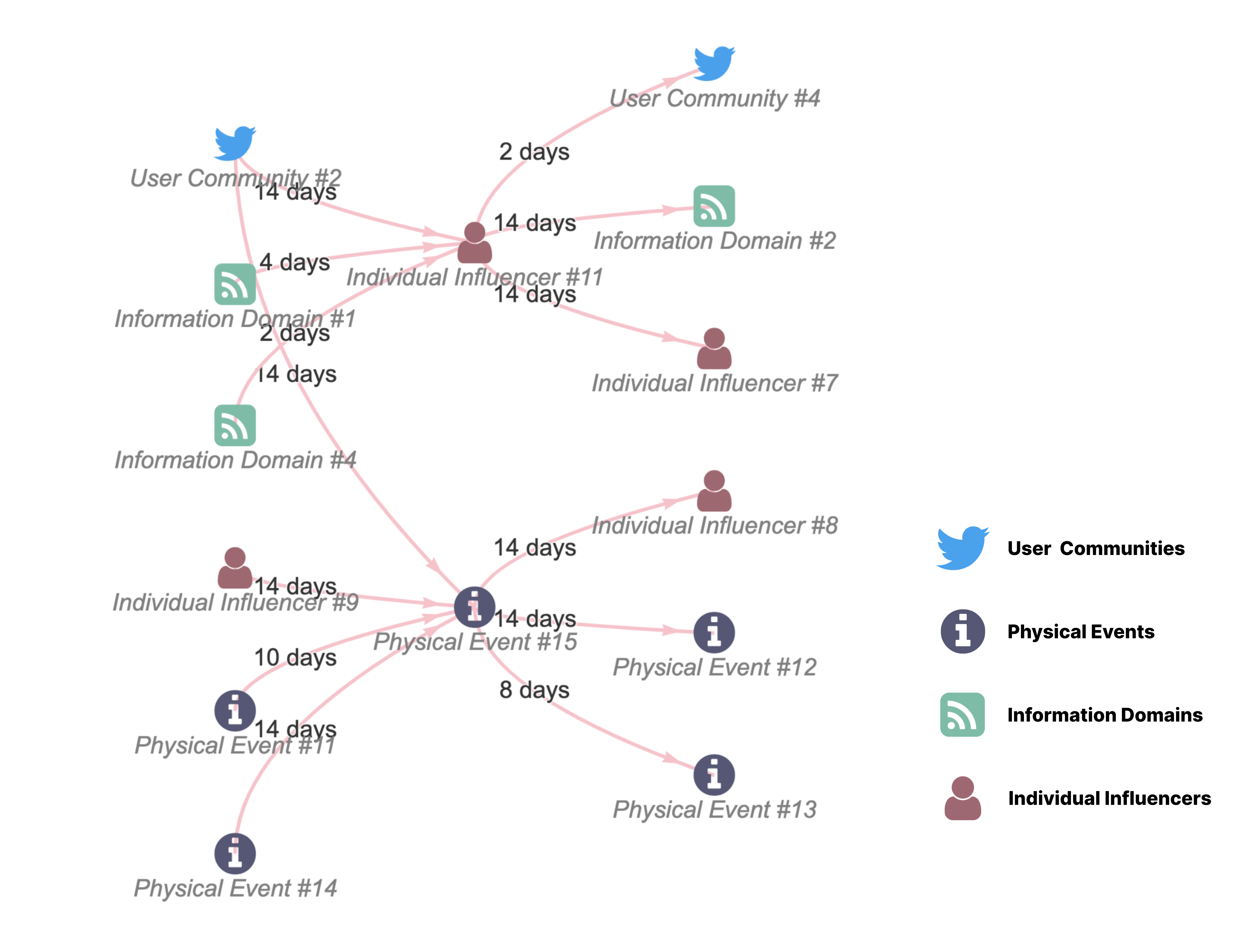}
    \caption{An illustrative example of a derived influence network.}
    \label{fig:task}
\end{figure}

Figure~\ref{fig:task} shows an example latent influence network (computed by the proposed algorithm), including the four types of entities we defined above.

\section{The Solution Pipeline}
\label{sec:solution}
To compute the latent influence network, we organize the set of media posts (provided as input) into a graph, compute ideological embeddings for graph nodes, decide on entities to consider, and finally test these entities for significant pairwise directional influence relations. The resulting set of uncovered directional influence links (together with the entities they link) then constitutes the produced latent influence network. The pairwise influence test itself is a plug-in module in the framework that determines whether or not directional influence is likely between a given pair of entities. For simplicity, below we shall use {\em lagged correlations in belief embedding\/} as an indicator of potential influence, implying that a potential influence link is suspected when the time-series of one entity can predict the time-series of another. This concept aligns with Granger causality~\cite{seth2007granger} but is not in itself a proof of causal relations. Future incarnations of the framework can replace this test with others that align better with actual  causality~\cite{pearl2009causality} (based on observations of belief embedding). More specifically, the execution pipeline consists of four stages: 

\begin{enumerate}
    \item {\it Interaction graph construction:\/} This stage involves organizing the input posts into a bipartite interaction graph whose nodes are (i) the individual users (who posted on the media), and (ii) their posts. 
    \item {\it Interaction graph cleaning:\/}
    Since we might not have visibility into all interactions, a missing-link prediction approach~\cite{npugraph} is used to predict (likely) missing links in the computed graph, as well as to eliminate spurious (i.e., unlikely) ones.
    \item {\it Dynamic ideological representation learning:\/} Next, we compute the unsupervised interpretable ideological embedding~\cite{li2022unsupervised} for all nodes in the (cleaned) interaction graph, mapping them (both users and posts) into an interpretable, low-dimensional, latent space. These lower-dimensional latent representations are computed for successive time intervals to capture the evolution of espoused beliefs over time.  
    \item {\it Influence pathway discovery:\/} This stage focuses on identifying meaningful entities and valuable patterns of influence propagation among those entities.
    A community detection toolkit is first used to group individual users in the interaction graph into {\em user community\/} entities. For each found community, the ideological embeddings of community members are averaged to yield a single community-wide time-series. Popular users are separately cast as potential {\em individual influencer\/} entities. Popularly cited URLs are similarly cast as {\em information domain\/} entities. Finally, {\em physical entities\/} are added based on supplied input. 
    Correlations are then computed between all entity time-series involved. Edges are plotted to depict large correlations. These edges (and the entities they connect) form the output latent influence network.
\end{enumerate}

\begin{figure}[t]
    \centering
    \includegraphics[width = 1.\columnwidth]{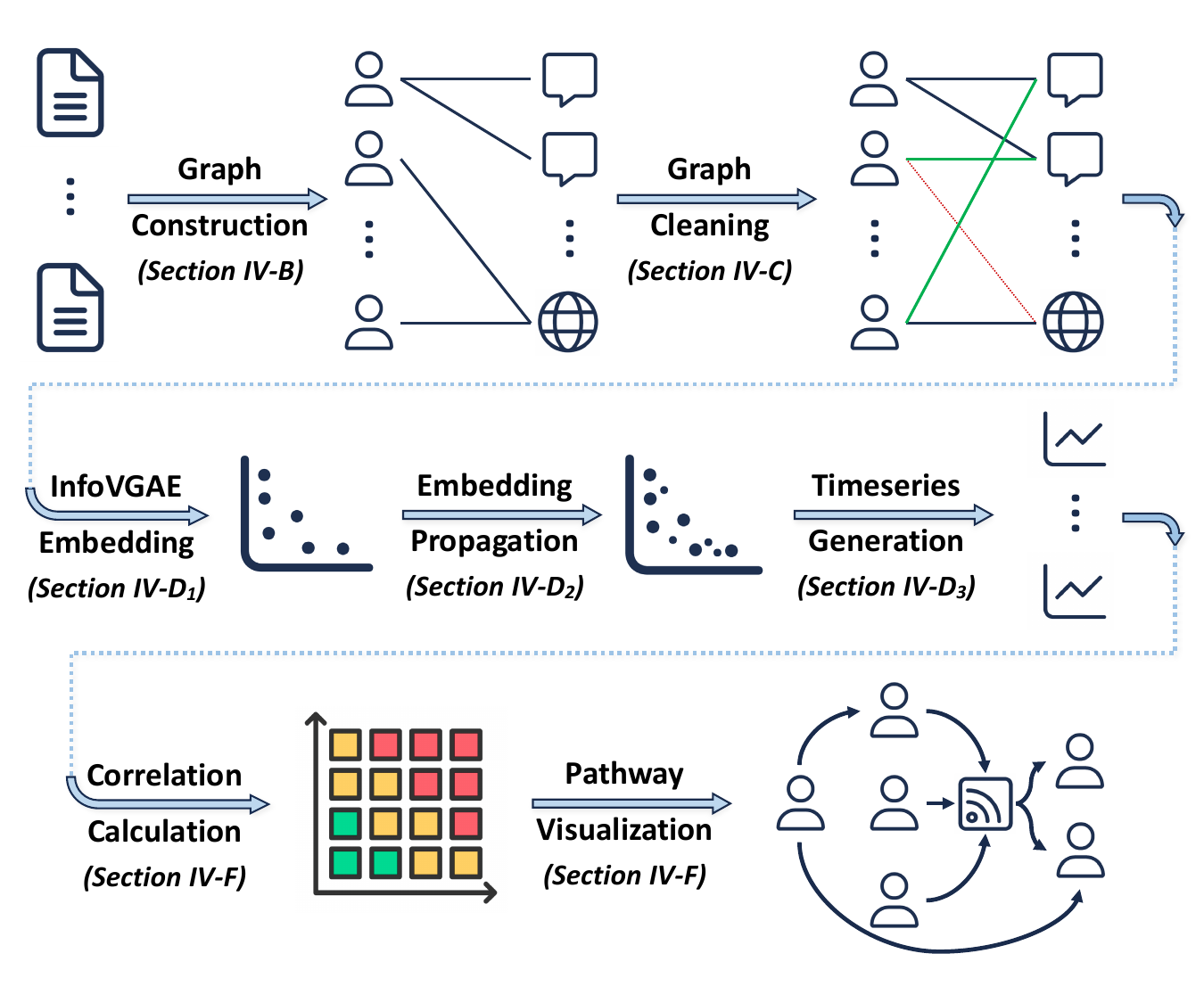}
    \caption{An overview of the proposed pipeline.}
    \label{fig:pipeline}
\end{figure}

\noindent
Figure~\ref{fig:pipeline} depicts the overall workflow. The four stages mentioned above are discussed in more detail in Section~\ref{sec:graph}, Section~\ref{sec:npugraph},  Section~\ref{sec:infovgae}, and Section~\ref{sec:community}, respectively. 

\subsection{Interaction Graph Construction}
\label{sec:graph}
The first step of the pipeline is to design a data structure that can model the relationship between the users and the content they post. The datasets collected for this paper are from Twitter (currently X), where posted tweets can be collected using API calls (subject to a rate limit) together with the posting users and timestamps. Based on the raw data, we extend the user-post graph by extracting embedded URLs in the posted tweets and treating them as separate nodes linked back to the posting users. The result is a bipartite graph of user-content relationships, where the two types of nodes are media users and posts (e.g., tweets and posted URLs). We call them {\em user nodes\/} and {\em assertion nodes\/}, respectively.


\subsection{Graph Cleaning: nPUGraph}
\label{sec:npugraph}
The initial interaction graph we create may suffer false positives (i.e., incorrectly attributed interactions) and false negatives (unobserved interactions). These instances of incorrect or missing interactions can significantly distort the subsequent process of learning ideology representations, resulting in the introduction of noticeable noise into the ideological embedding time-series data. To address this issue, we incorporate the nPUGraph method~\cite{npugraph}, previously introduced in our research, as a data-cleaning step. This method effectively addresses both the prediction of missing interactions and the removal of erroneous ones, thereby enhancing the overall quality of the interaction graph.

\subsection{Dynamic Ideological Representation Learning}
\label{sec:infovgae}
\begin{figure}
    \centering
    \includegraphics[width = 1.\columnwidth]{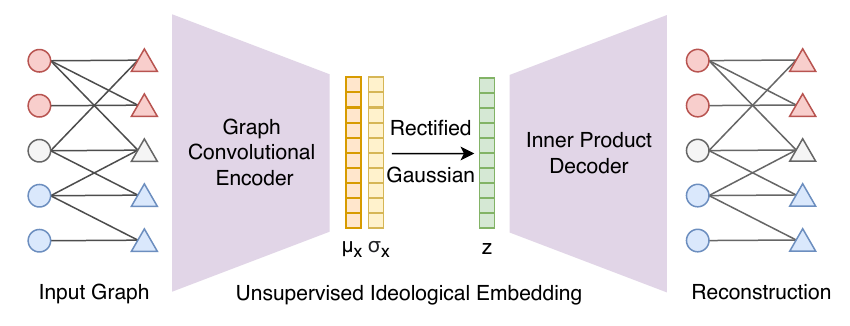}
    \caption{The InfoVGAE architecture. }
    \label{fig:infovgae}
\end{figure}
We map the users and assertions into an explainable latent space using InfoVGAE~\cite{li2022unsupervised}, shown in Figure~\ref{fig:infovgae}. InfoVGAE is a variational graph auto-encoder that projects inputs into the positive quadrant only (of the latent space). Its
overall loss function is further modified to ensure orthogonality of projection coordinates, in addition to optimizing the evidence lower bound (ELBO) of the typical VAE. Prior work has shown that these changes lead to interpretable embedding properties, as the orthogonality constraints (combined with restrictions to the positive quadrant) cause different ideological echo-chambers to be mapped onto orthogonal axes, while users and posts at the intersection of the echo-chambers (i.e., those who are more neutral) are mapped closer to the origin~\cite{li2022unsupervised}. 

The above characteristics have important consequences. They imply that the coordinate value of an entity's embedding on a given axis has an interpretable meaning. When that value decreases, the entity must be expressing more neutral beliefs with respect to the ideology represented by that axis. Similarly, when the coordinate value increases, the entity must be getting more entrenched in the ideology represented by the axis.
Said differently, changes towards higher coordinate values on an axis represent ideological rhetoric escalation (for the corresponding ideology). Similarly, changes towards lower coordinate values on the axis represent rhetoric de-escalation (for the corresponding ideology). 

Unfortunately, the space complexity of InfoVGAE is of the order of the product of users and assertions. To make it scale, we use InfoVGAE to compute the embedding for popular users and assertions only, then propagate the computed node embeddings to their neighbors in the bipartite user-assertion graph, such that previously excluded less popular nodes inherit the average of their neighbors with known ideology embeddings. The process is repeated in successive time windows.  


\begin{table*}[t]
\caption{Dataset Statistics.}
\centering
\begin{tabular}{c|c|c|c|c}
\toprule
\multicolumn{1}{l}{} & \#Users & \#Messages & \#Assertions & Date Range \\ \hline
2022 French Election & 389,187 & 3,195,579 & 829,484 & 02/15/2022-04/11-2022 \\ \hline
Philippine & 204,320 & 354,370 & 220,169 & 01/01/2023-06/28/2023 \\ \hline
Russophobia & 103,917 & 243,088 & 101,057 & 05/01/2022-04/15/2023 \\ \bottomrule
\end{tabular}%
\label{table:dataset_stats}
\end{table*}

\begin{figure*}[t]
    \centering
    \begin{subfigure}{0.32\textwidth}
        \includegraphics[width = \textwidth]{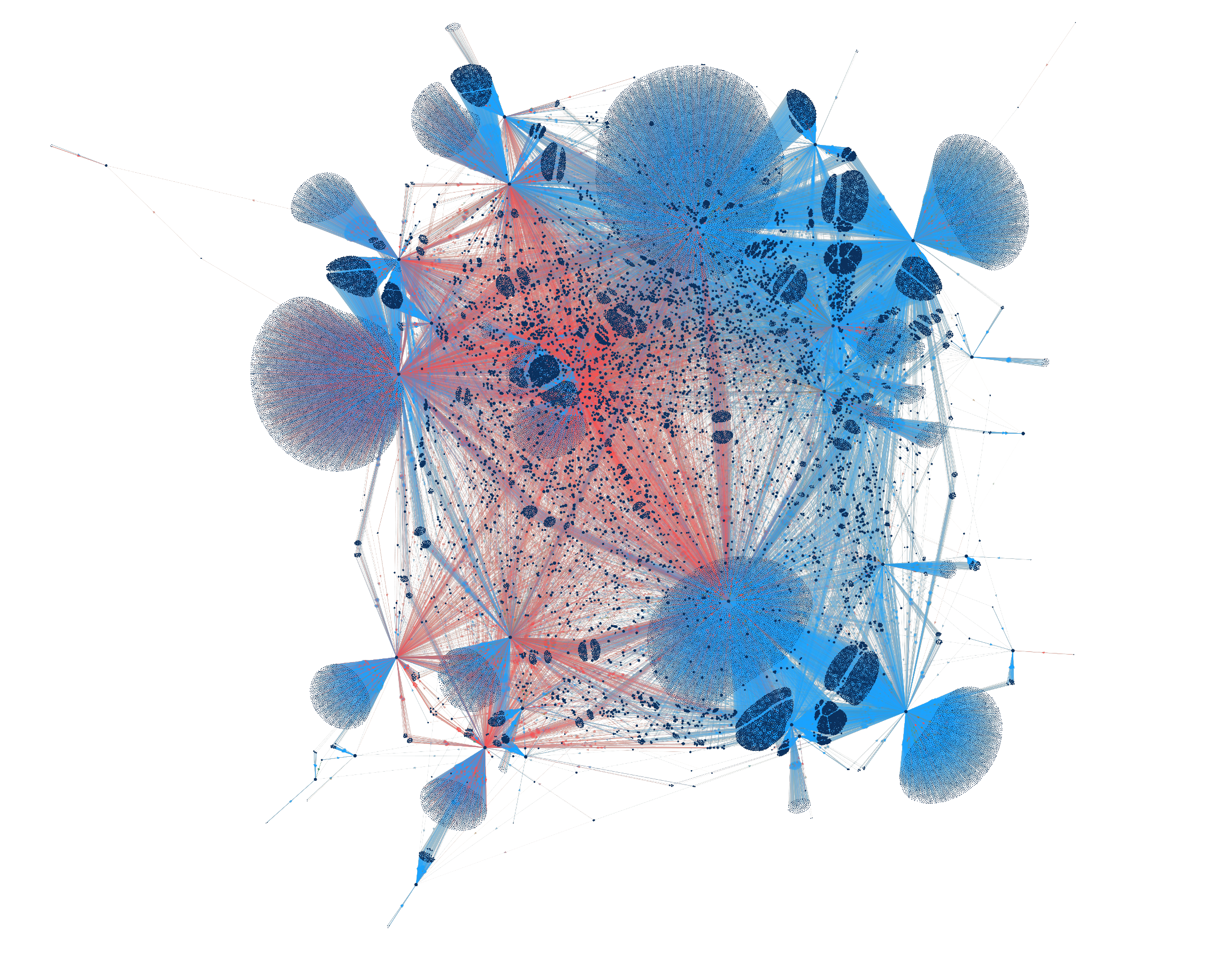}
        \subcaption{The French election dataset user graph.}
        \label{fig:graphvis_1b}
    \end{subfigure}
    \begin{subfigure}{0.31\textwidth}
        \includegraphics[width = \textwidth]{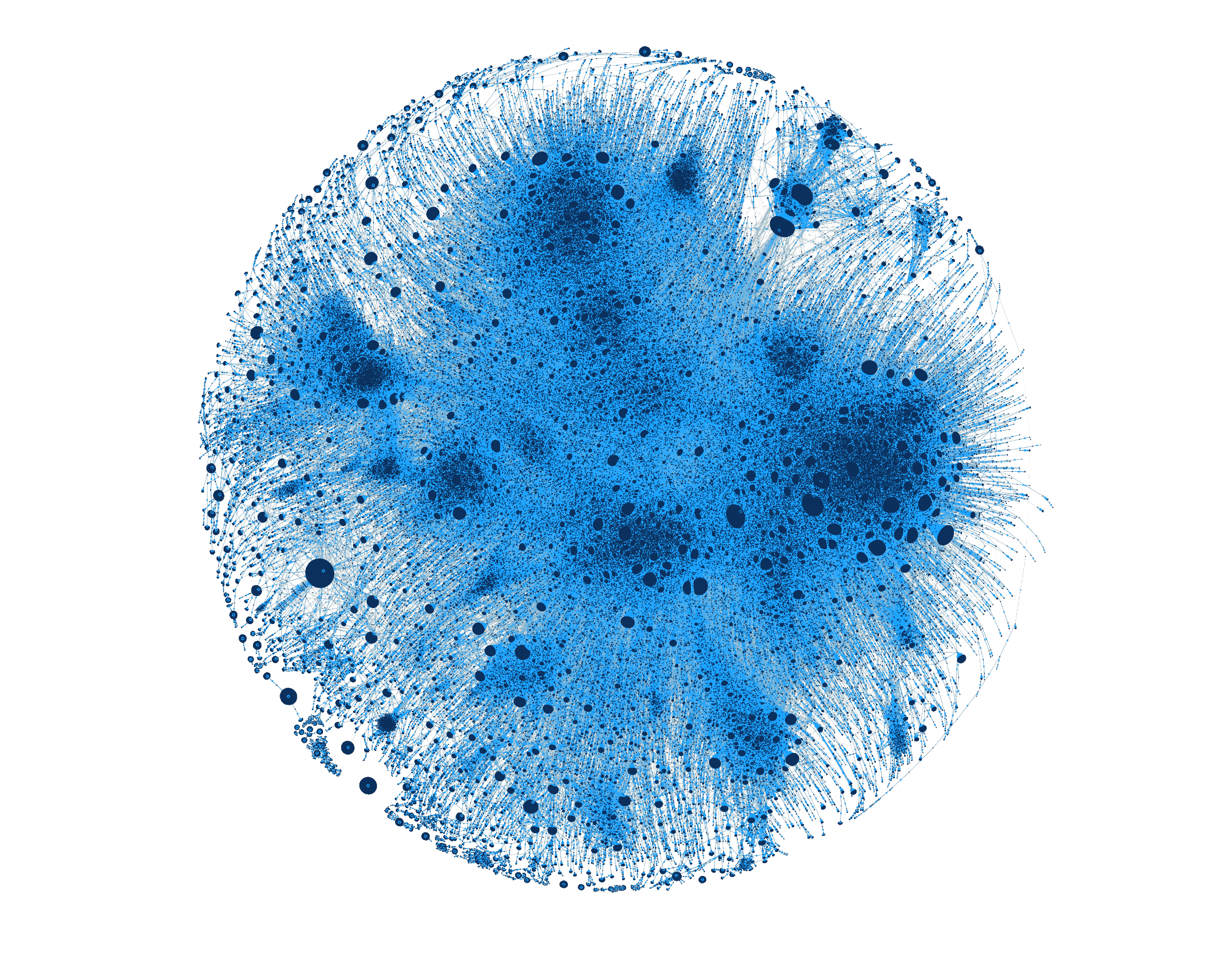}
        \subcaption{The Philippines dataset user graph.}
        \label{fig:graphvis_2a}
    \end{subfigure}
    \begin{subfigure}{0.31\textwidth}
        \includegraphics[width = \textwidth]{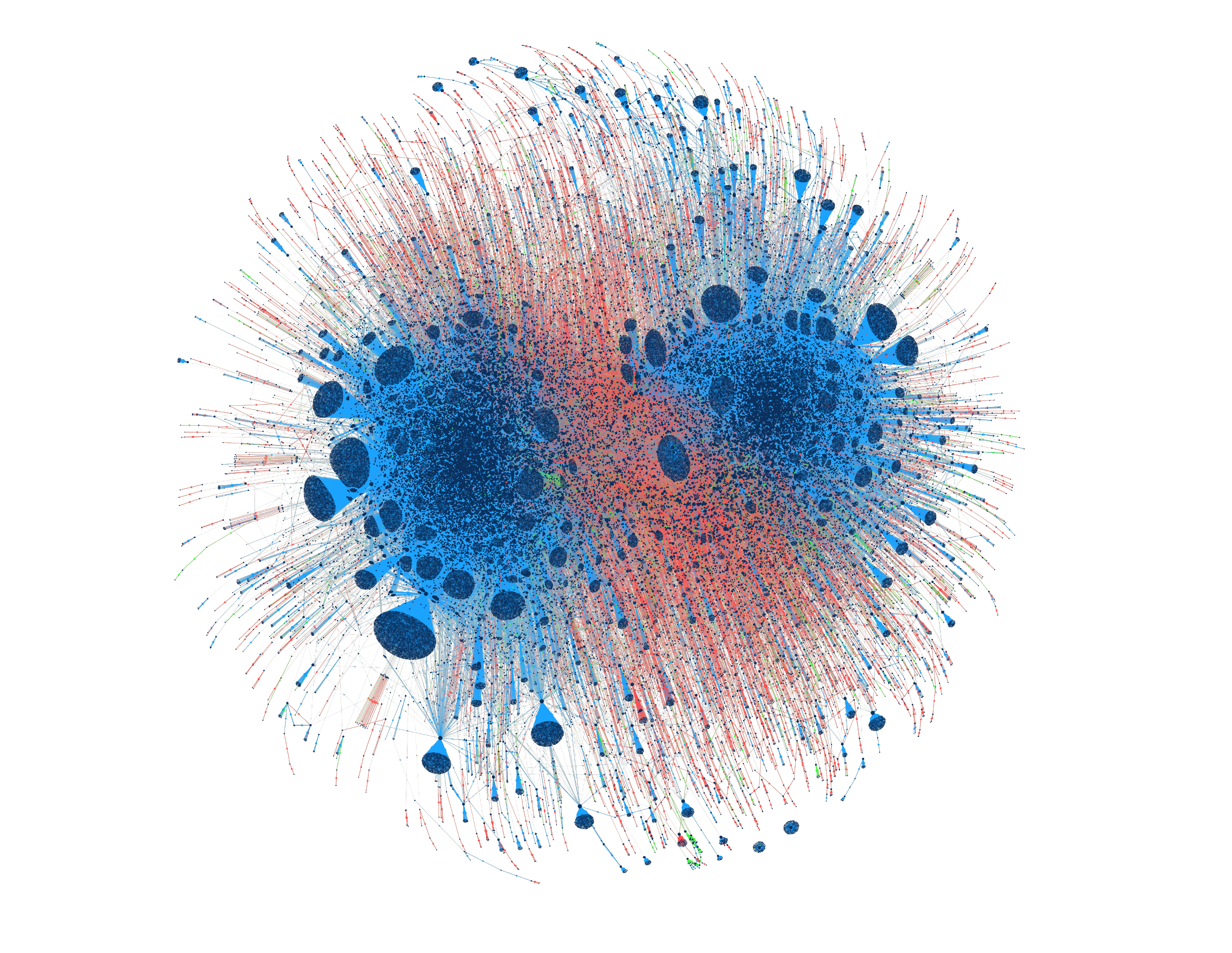}
        \subcaption{The Russophobia dataset user graph.}
        \label{fig:graphvis_russo}
    \end{subfigure}
    \caption{User graph visualizations. Nodes are colored dark blue. Repost (retweet) edges are colored blue, reply edges are colored red, and quotation edges are colored green.}
    \label{fig:graphvis}
\end{figure*}

\begin{table*}[t]
\caption{Algorithm Settings.}
\centering
\begin{tabular}{c|c|c|c|c}
\toprule
\multicolumn{1}{l}{} & Window Length & Time Shift & Time Lag Threshold & Correlation Threshold \\ \hline
2022 French Election & 20 days & 1 day & 5 days & 0.7 \\ \hline
Philippine & 20 days & 2 days & 5 days & 0.5\\ \hline
Russophobia & 20 days & 2 days & 5 days & 0.4 \\ \bottomrule
\end{tabular}%
\label{table:dataset_settings}
\end{table*}

\begin{figure*}[t]
    \centering
    \includegraphics[width = 0.92\textwidth]{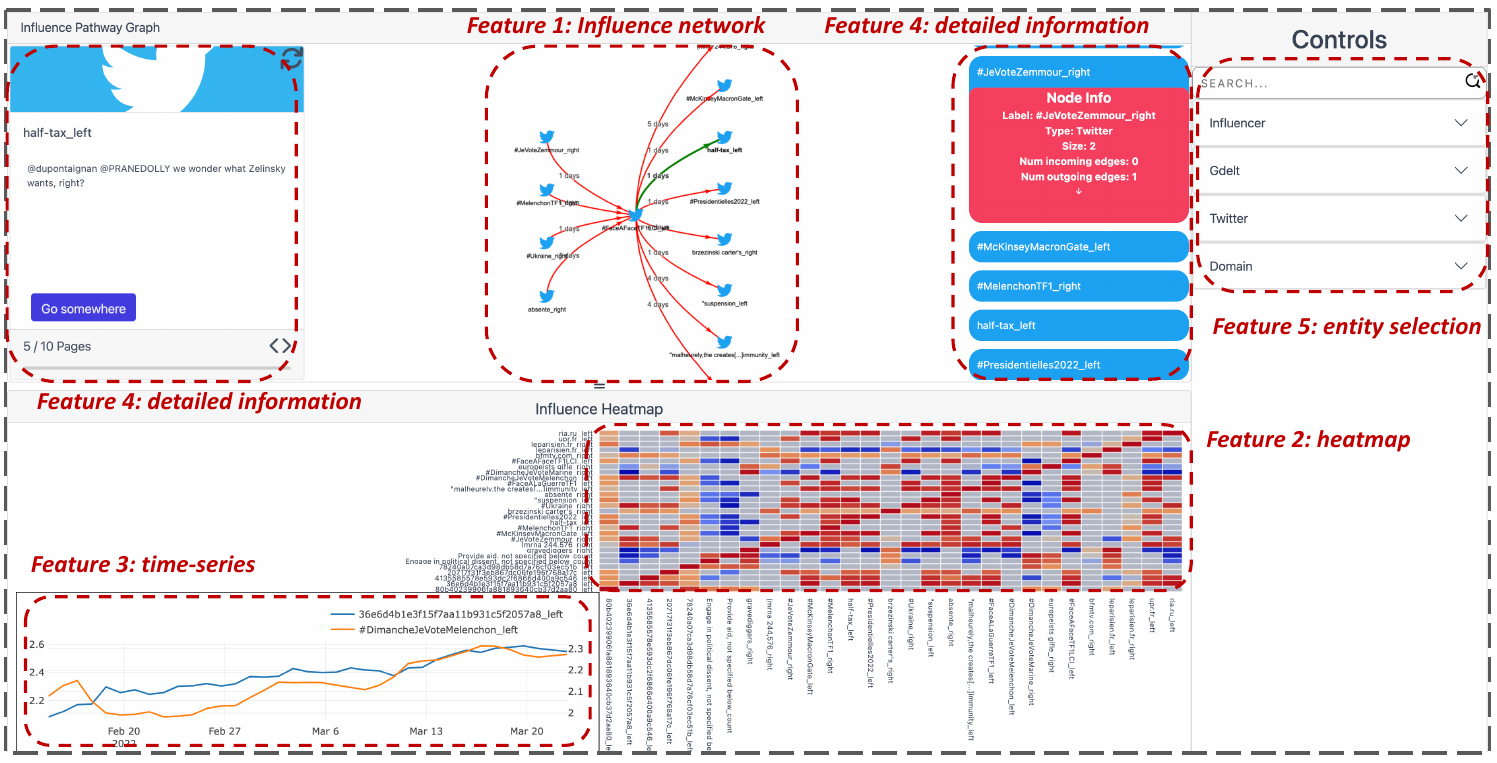}
    \caption{An overview of the user interface. The service can be publicly accessed via https://incas.csl.illinois.edu/tasks.}
    \label{fig:development}
\end{figure*}

\subsection{Influence Pathway Discovery}
\label{sec:community}
After the ideological embedding is computed for all users and assertions, we identify the important entities to be included in our influence network. As mentioned earlier, we include as entities all nodes corresponding to popular referenced URLs (information domain entities) and all popular users (individual influencer entities), then lump the remaining users into community entities, computing an average embedding time-series for each.
We then introduce physical entities. For a proof of concept, we consider events from GDELT, an external database of the world's real-life events. Among 311 event types that GDELT provides, we selected 15 that are relevant to political activity, such as ``provide economic aid", ``investigate military action", and ``obstruct passage".  For each such event type, we produce a time series of event counts versus time. 
With candidate entities selected as discussed above, we calculate a lagged Pearson correlation for each entity pair. 
If a spike occurs at a given offset, we say that an influence edge was found (in the direction from the leading time-series to the lagging one). The identification of all such (high-correlation) edges completes graph construction.

\section{Case Studies}
\label{sec:evaluation}
Next, we offer a brief example of using this approach on social media data sets collected on recent events. 

\subsection{Datasets}

We test our pipeline on three datasets collected from X (formerly Twitter). They are (i) the 2022 French election, (ii) recent geopolitical events in the Philippines, and (iii) a Russian influence campaign aiming to garner sympathy from claims of ``Western Russophobia". Some basic statistics for each are shown in Table~\ref{table:dataset_stats}. The corresponding user interaction graphs are displayed in Figure~\ref{fig:graphvis}.

\begin{enumerate}
    \item \textbf{2022 French Election:} The dataset focuses on the multiple presidential candidates, the relevant media outlets, and the candidates' supporters during the time leading to the 2022 French presidential election. Figure~\ref{fig:graphvis_1b} shows the corresponding user interaction graph, where one can distinguish several clusters. They happen to correspond to the different candidates and their followers.
    \item \textbf{Philippines:} The dataset covers a multitude of topics from the Philippines' geopolitical landscape from the first half of the year 2023. User interactions are shown in Figure~\ref{fig:graphvis_2a}.
    \item \textbf{Russophobia:} The dataset covers posts on a key talking point of pro-Russian messaging during the Russia-Ukraine war, namely, the claims of \textit{``Western Russophobia"} (referring to the allegation of unfair/hostile world attitudes towards citizens of Russia), collected for nearly a year starting 05/01/2022. Posts reacting to this line of messaging are also included. Figure~\ref{fig:graphvis_russo} shows the user interaction graph, where one can clearly distinguish two groups of users. They happen to be those siding with the pro-Russia messaging and those against it.
\end{enumerate}

Table~\ref{table:dataset_settings} reports empirically tuned parameter settings for optimal performance on each dataset, including {\em Window Length\/} (the interval of time, in days, over which each ideological embedding value is computed), {\em Time Shift\/} (how much to slide this window by in each step for time-series generation), {\em Time Lag Threshold\/} (the maximum time lag we consider between two time-series in searching for correlations), and {\em Correlation Threshold\/} (the minimum Pearson-correlation used to identify an influence edge).

\subsection{Selected Observations}
\begin{figure*}[t]
    \centering
    \begin{subfigure}[b]{0.49\linewidth}
    \centering
    \includegraphics[width = \linewidth]{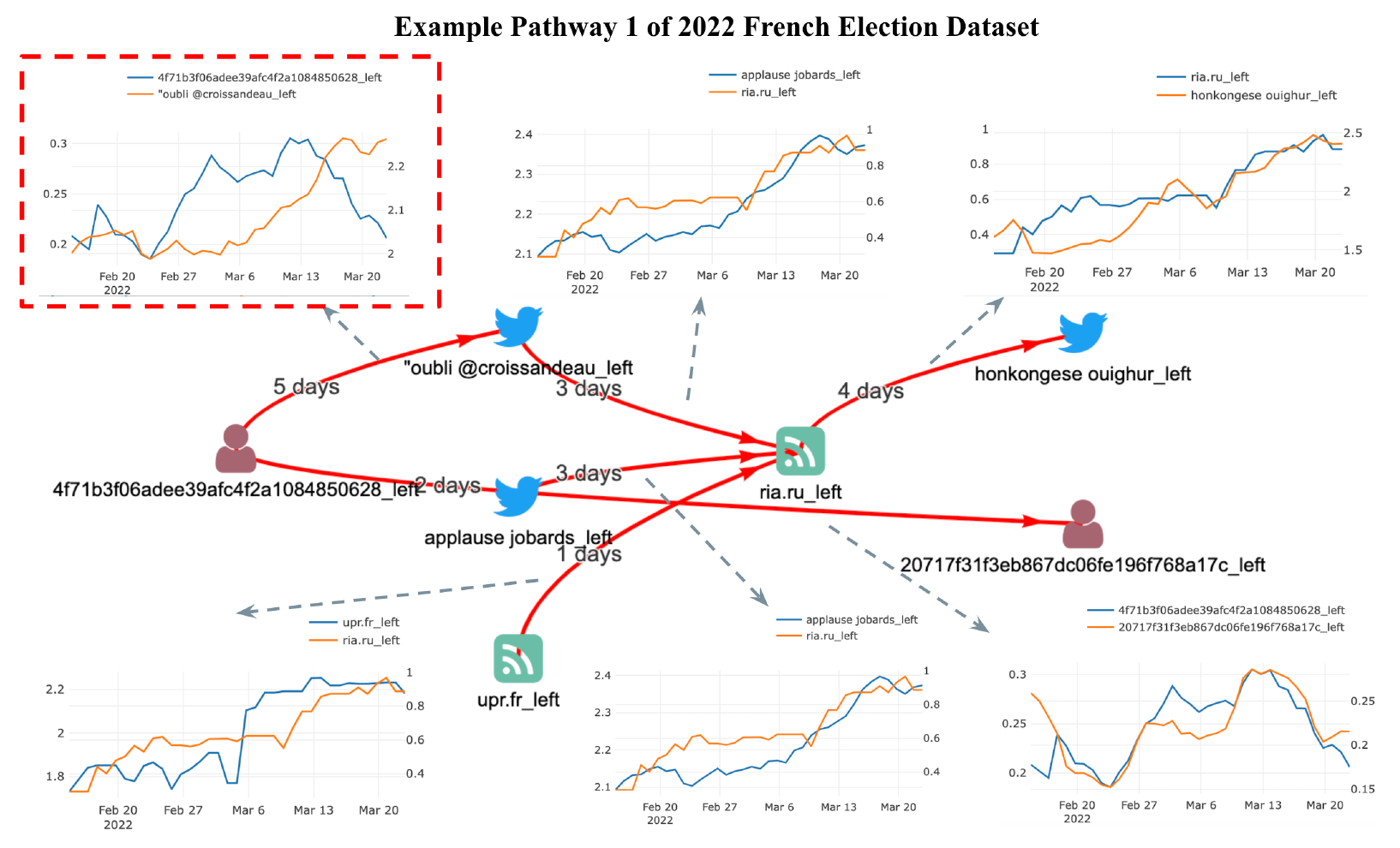}
    \caption{Example pathway 1 from the French election dataset.}
    \label{figure:influence_pathways_french1}
    \end{subfigure}
    \begin{subfigure}[b]{0.49\linewidth}
    \centering
    \includegraphics[width = \linewidth]{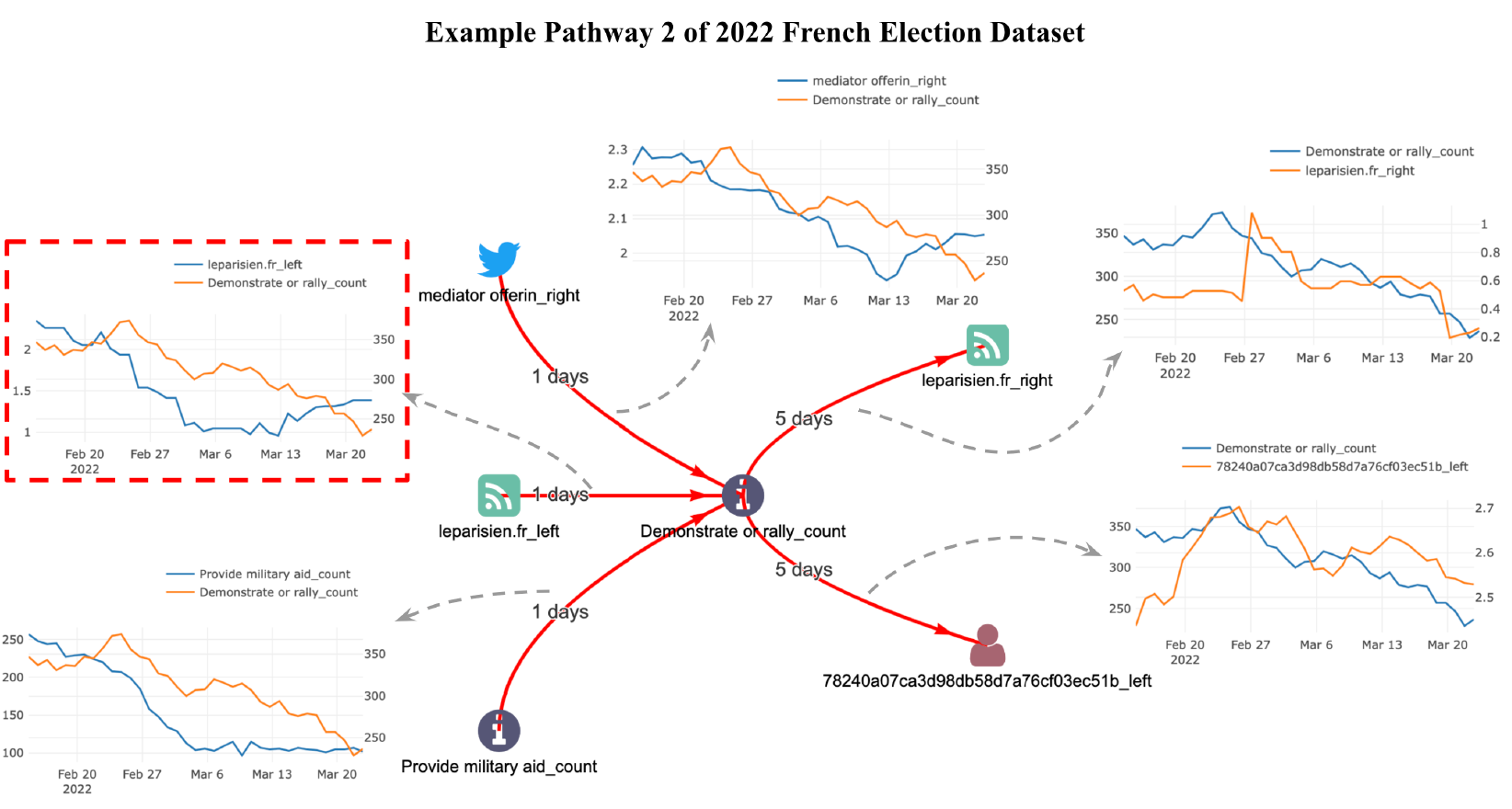}
    \vspace{0.2cm}
    \caption{Example pathway 2 from the French election dataset.}
    \label{figure:influence_pathways_french2}
    \end{subfigure}
    \begin{subfigure}[b]{0.49\linewidth}
    \centering
    \includegraphics[width = \linewidth]{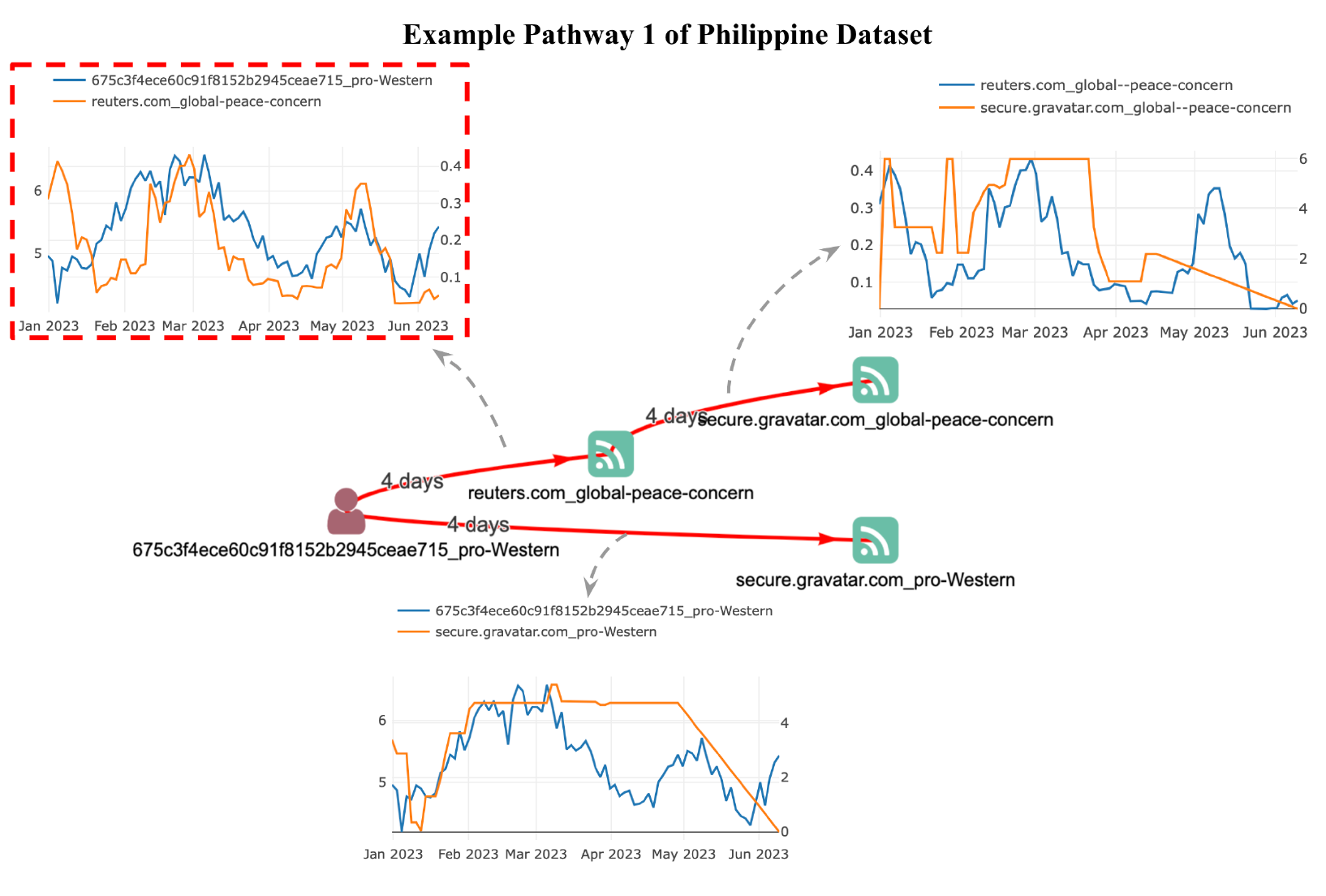}
    \caption{Example pathway from the Philippines dataset.}
    \label{figure:influence_pathways_philippine}
    \end{subfigure}
    \begin{subfigure}[b]{0.49\linewidth}
    \centering
    \includegraphics[width = \linewidth]{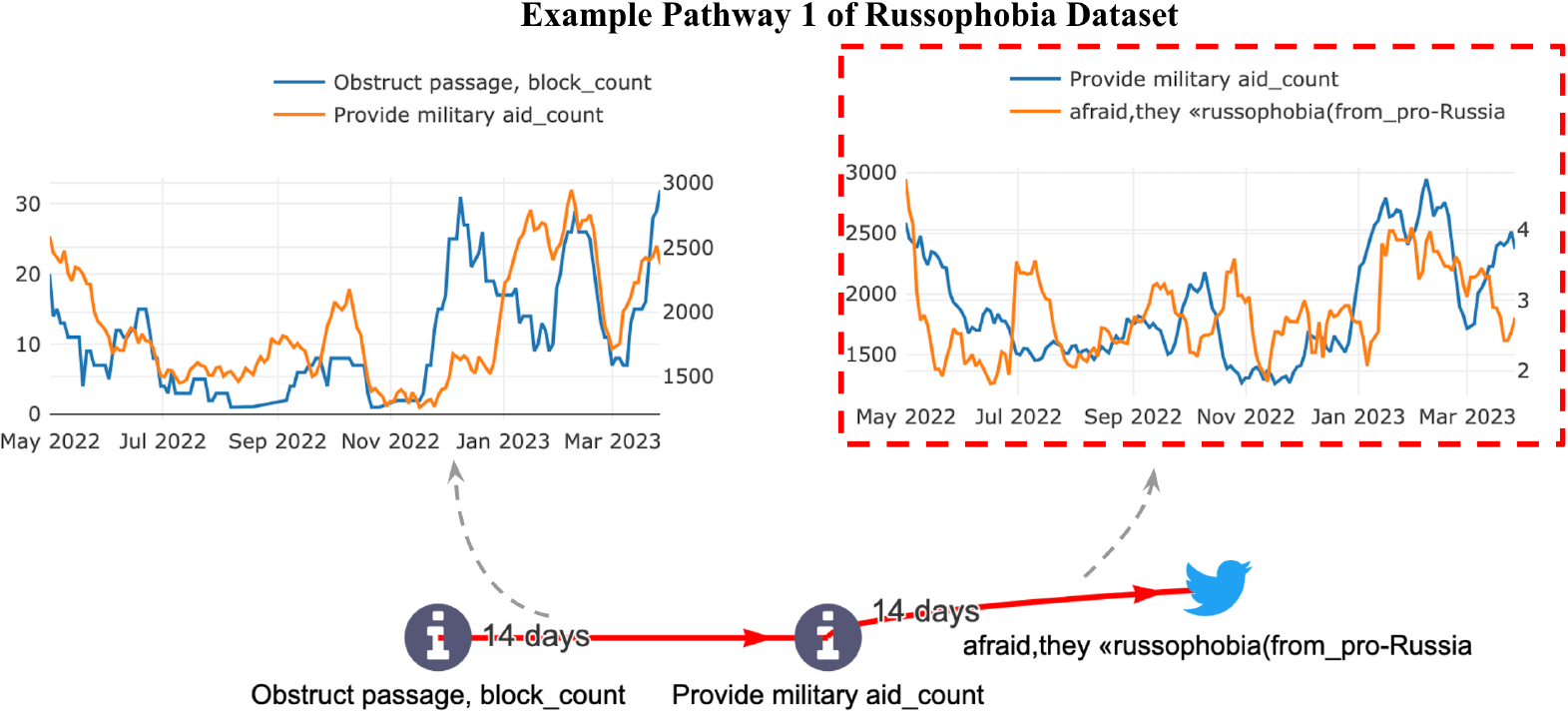}
    \vspace{0.8cm}
    \caption{Example pathway from the Russophobia dataset.}
    \label{figure:influence_pathways_russo}
    \end{subfigure}
    \caption{Examples of ideology influence pathways.}
    \label{figure:case}
\end{figure*}

Figure~\ref{fig:development} shows the interface used to inspect discovered influence pathways. The screen allows the operator to inspect parts of the influence network, view a heatmap of identified (lagged) correlations between key entities, drill down into the involved entity-pair's ideological embedding time-series for any cell in the heatmap, list the specific posts that explain the underlying entity-espoused beliefs (for influencers, domains, and communities), and control entity selection to view specific parts of the overall influence graph. Example snippets of the computed influence graphs are shown in Figure~\ref{figure:case}.

It is not the purpose of this paper to report political findings. Thus, we obscure the names of the entities depicted and focus on the general interpretations of the discovered pathway examples. 
Following the influence graphs shown Figure~\ref{figure:case} (in the direction of the arrows, left-to-right), Figure~\ref{figure:influence_pathways_french1} shows a key influencer in the French election, and identifies two Twitter communities influenced by them, as well as a news portal that correlates with their view. Other links are also shown, such as a downstream community influenced by the news portal.
Figure~\ref{figure:influence_pathways_french2} shows another example from the French election, where the rhetoric of a news portal and an online community seem to predict protest counts. The graph also shows other entities apparently influenced by the protests. 
Figure~\ref{figure:influence_pathways_philippine} shows the impact of a key influencer in the Philippines dataset, as well as multiple news media that follow their view, including the reuters.com domain. 
Finally, 
Figure~\ref{figure:influence_pathways_russo} shows a pathway from the Russophobia dataset, depicting how a series of GDELT events, including military aid (to Ukraine) incite increased Russophibia rhetoric within a downstream Twitter community. 

These examples offer a flavor of what the pathway discovery analysis can support. The actual results can be fine-tuned by controlling the selection of entities to consider. Such control may be exerted either directly (e.g., by selecting the appropriate physical events, portals, or influencers to include) or indirectly (e.g., by selecting the population segmentation algorithm responsible for generating the community entities of interest).

\section{Discussion and Future Work}
\label{sec:discussion}
This influence pathway discovery solution presented in this paper opens up opportunities for future expansion.  It is meant to offer a flexible framework where individual modules can be replaced and enhanced over time. Some such opportunities are discussed below. The first desirable improvement lies in replacing the lagged correlation approach (inspired by Granger causality) with alternative methods that more accurately capture genuine causal effects. Identified (ideally, causal) links can further be viewed as dynamical stimulus/response systems modeled by transfer functions that capture the (stimulus/response) relation between their input and output time-series. An advantage of such models is that they can capture (possibly nonlinear) cumulative effects (where a response is a function of the {\em integral\/} of the stimulus) as well as novelty effects (where a response is proportional to the {\em change\/} in stimulus), among other functions that better model the dynamics of social belief adoption. 

The framework also offers the flexibility to integrate new types of entities for analysis. For example, new types of user communities may be identified by alternative community detection algorithms, and new physical entity types can be integrated besides GDELT events. The community and event names can be automatically generated using large language models (LLMs), applied to summarize or label sets of community posts. 
More substantial extensions are also possible, such as what-if analysis. We can simulate how influence might spread among entities when newly generated messages or information campaigns are introduced into the social medium. 

Finally, scalability remains a challenge. Our ideological embedding and community detection methods do tackle scalability concerns, allowing us to directly analyze extensive social graphs comprising millions of nodes. Next, streaming versions of these solutions are desired to allow continuous updates as new social media posts arrive. We hope the above framework will ultimately contribute to the mitigation of misuse in the information space.
\section{Conclusions}
\label{sec:conclusion}

We presented a novel influence graph discovery pipeline that differs from existing methods in its reliance on ideological embedding to uncover evidence of influence. It defines influence as the act of impacting human beliefs and infers changes in beliefs by mapping observed posts to an appropriately-constructed interpretable latent space. 
An interactive user interface and the corresponding backend were also implemented that enable us to analyze real-world data sets. Case studies were conducted to demonstrate the functionality of the proposed approach. The work serves as a proof of concept for influence network tomography based on belief embedding. 

\section*{Acknowledgements}
Research reported in this paper was sponsored in part by DARPA award HR001121C0165, DARPA award HR00112290105, and DoD Basic Research Office award HQ00342110002. It was also supported in part by ACE, one of the seven centers
in JUMP 2.0, a Semiconductor Research Corporation (SRC)
program sponsored by DARPA. 

\bibliographystyle{IEEEtran}
\bibliography{main}

\end{document}